\documentclass[twocolumn]{elsart}
\usepackage{natbib}
\usepackage{graphicx}
\usepackage{amssymb}

\begin{document}

\begin{frontmatter}

\title{Critical Review of Basic Afterglow Concepts}
\author[label1,label2]{Jonathan Granot}
\ead{j.granot@herts.ac.uk}
\ead[url]{http://star.herts.ac.uk/$\sim$granot}

\address[label1]{Kavli Institute for Particle Astrophysics and Cosmology,
Stanford University, P.O. Box 20450, MS 29, Stanford, CA 94309, USA}
\address[label2]{Centre for Astrophysics Research, University of
Hertfordshire, College Lane, Hatfield, Herts, AL10 9AB, UK}

\begin{abstract}

The long lived afterglow emission that follows gamma-ray bursts (GRBs)
was predicted prior to its detection in 1997, in the X-rays, optical
and radio. It is thought to arise from the shock that is driven into
the external medium as the latter decelerates the relativistic outflow
that drives the GRB, and persists well after most of the energy in the
outflow is transferred to the shocked external medium. As the blast
wave decelerates, the typical emission frequency shifts to longer
wavelength. Recent observations following the launch of the {\it
Swift} satellite challenge the traditional afterglow modeling and call
into questions some of the basic underlying concepts. This brief
review outlines some of the major strengths and weaknesses of the
standard afterglow model, as well as some of the challenges that it
faces in explaining recent data, and potential directions for future
study that may eventually help overcome some of the current
difficulties.

\end{abstract}


\end{frontmatter}

\section{Introduction}
\label{ec:intro}

Gamma-ray bursts (GRBs) are thought to originate from an
ultra-relativistic outflow that is driven by a sudden catastrophic
energy release from a compact stellar central source. The prompt
$\gamma$-ray emission is usually attributed to energy dissipation
within the relativistic outflow. The outflow sweeps up the external
medium and drives a strong relativistic shock into it, until the
outflow is eventually decelerated significantly and most of the energy
in the flow is transferred to the shocked external medium \citep[for
recent reviews see][]{Piran05,Meszaros06}. The shocked external medium
is thought to produce the long lived afterglow emission that is
detected in the X-rays, optical, and radio for days, weeks, and
months, respectively, after the GRB.

Afterglow emission was predicted \citep{PR93,Katz94,MR97,SP97} prior
to its detection in 1997 in the X-ray \citep{Costa97}, optical
\citep{vanParadijs97}, and radio \citep{Frail97}. Furthermore, its
basic observational properties (light curve and spectrum) initially
agreed rather well with the simple model predictions
\citep{Waxman97,WRM97}. This gave some credence to the basic picture
where the afterglow emission arises from the external shock going into
the ambient medium, where the typical emission frequency gradually
shifts to longer wavelength as the blast wave decelerates. 

As afterglow observations improved, however, new and more complicated
behavior was observed, which could not be accommodated by the simplest
version of the external shock afterglow model (which features a
spherical adiabatic blast wave propagating into a uniform external
medium). This called for new ingredients in the model, i.e. variants
of the simplest model. Some of these variants and their observational
signatures were predicted before they were observed, which again
provided support for the basic model.

The Launch of the {\it Swift} satellite in November 2004 marked a new
era in GRB research. Its ability to rapidly and autonomously slew to
the direction of GRBs that it detects and observe them in the X-rays,
UV, and optical has significantly improved our knowledge of the early
afterglow, from tens of seconds to hours after the onset of the GRB.
The complex behavior that was revealed challenges the traditional
afterglow theory, and even puts to question the basic underlying
picture. 

In \S ~\ref{sec:theory} a short overview is given of the standard
afterglow model, starting from the dynamics in the simplest case
(\S~\ref{sec:spherical}), continuing with the emission mechanism
(\S~\ref{sec:emission}), and finishing with variants of the simplest
model (\S~\ref{sec:variants}). Next, \S~\ref{sec:weak} outlines some
of the weaknesses of the standard afterglow model. In
\S~\ref{sec:Swift} a brief description is given of the major new
observations in the {\it Swift} era, stressing the difficulties and
challenges that they present for the standard afterglow
model. Finally, \S~\ref{sec:future} describes the crisis for afterglow
theory that was caused by the new observations, and discusses some
ideas for resolving the crisis as well as possible directions for the
future.

\section{Brief Overview of Standard Afterglow Theory}
\label{sec:theory}

\subsection{The Simplest Hydrodynamic Model: a ``Spherical Cow''}
\label{sec:spherical}

The simplest version of the standard afterglow model features a
spherical relativistic adiabatic (with negligible energy losses or
gains) blast wave that propagates into a uniform external medium, of
mass density $\rho_{\rm ext} = n_{\rm ext}m_p$. Initially, the outflow
is decelerated by a reverse shock\footnote{The shocked outflow as a
whole is further decelerated by work that is performed on it by the
shocked external medium across the contact discontinuity that
separates them, while each of its fluid elements is decelerated by the
pressure gradients that form within it.}, while the external medium is
swept-up by the forward shock, where the two shocked regions are
separated by a contact discontinuity. The ejecta are decelerated by
the reverse shock from their original Lorentz factor, $\Gamma_{\rm
ej}$, to a smaller Lorentz factor, $\eta$, where $\Gamma_{\rm ej} >
\eta \gg 1$. 

If $\Gamma_{\rm ej}^2\rho_{\rm ext}/\rho_{ej} \gg 1$, where $\rho_{\rm
ej}$ and $\rho_{\rm ext}$ are the proper mass densities (more
generally, one should use the ratio of enthalpy densities) of the
original ejecta and the external medium, respectively, then the
reverse shock is relativistic. In this case the relative Lorentz
factor between the upstream and downstream fluids across the reverse
shock, $\Gamma_{\rm RS} \approx (\Gamma_{\rm ej}/\eta+\eta/\Gamma_{\rm
ej})/2$, is given by $\Gamma_{\rm RS} \approx (\Gamma_{\rm
ej}/2)^{1/2}(\rho_{\rm ext}/\rho_{\rm ej})^{1/4} \gg 1$. 
If $\Gamma_{\rm ej}^2\rho_{\rm ext}/\rho_{ej} \ll 1$ then the reverse
shock is Newtonian and $\eta \approx\Gamma_{\rm ej}$ while $\Gamma_{\rm
RS}-1 \approx (4/7)\Gamma_{\rm ej}^2(\rho_{\rm ext}/\rho_{\rm ej}) \ll
1$ \citep{SP95}. This basic result may be obtained by equating the ram
pressure (momentum flux) of the incoming fluids (ejecta and external
medium) in the rest frame of the shocked fluids, $\rho_{\rm
ej}(\Gamma_{\rm RS}^2-1) \approx \rho_{\rm ext}\eta^2$. 

If the reverse shock is initially Newtonian, then a reasonable spread
in the Lorentz factor of the outflow ($\Delta \Gamma_{\rm ej} \sim
\Gamma_{\rm ej}$) would cause the reverse shock to strengthen more 
rapidly as it crosses the ejecta shell (since the shell starts to
spread in the radial direction causing a faster drop in its density)
in such a way that it becomes mildly relativistic when the reverse
shock finishes crossing the shell. Therefore, in this case as well (as
for an initially relativistic reverse shock), most of the energy is
transferred to the shocked external medium within a single shell
crossing by the reverse shock.

\begin{figure}
\includegraphics[width=1.0\columnwidth]{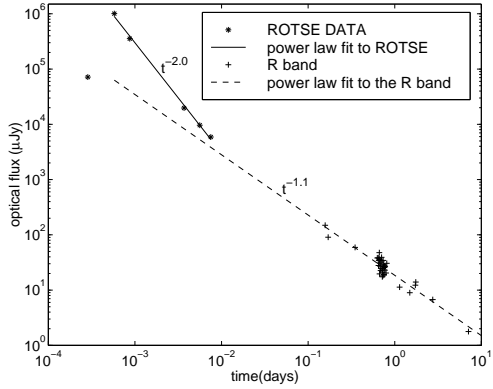}
\caption{Optical light curve of GRB~990123, where the early Robotic
Optical Transient Search Experiment (ROTSE) measurements (the
``optical flash''), excluding the first one, and the later R-band data
are each described by a different power law, and attributed to the
reverse shock and to the forward shock, respectively
\citep[from][]{SP99b}.}
\label{fig:optical-flash}
\end{figure}

A bright optical emission, referred to as an ``optical flash'' was
predicted from the reverse shock on a time scale similar to that of
the prompt GRB emission \citep{SP99a}. Soon thereafter it was observed
in GRB~990123 \citep{Akerlof99} and found to be in good agreement with
the theoretical expectations \citep[][see
Figure~\ref{fig:optical-flash}]{SP99b,MR99}. It was also supported by
the detection of a ``radio flare'' \citep{Kulkarni99}, which was
attributed to the emission from the shocked ejecta well after the
passage of the reverse shock, as its electrons cool adiabatically at
the back of the self-similar hydrodynamic profile and their typical
emission frequency passes through the radio after a day or so. This
was considered another success of what has by then become the standard
afterglow model.

After the initial stage in which the original outflow is significantly
decelerated, most of the energy is given to the shocked external
medium and the flow becomes self-similar \citep{BM76}. The
deceleration epoch ends and the self-similar stage starts at a radius
$R_{\rm dec} \approx (3E/4\pi\rho_{\rm ext}c^2\eta^2)^{1/3}$ and an
observed time $t_{\rm dec} \approx R_{\rm dec}/2c\eta^2$, where $\eta
\approx \min\left[\Gamma_{\rm ej},(\Gamma_{\rm ej}/2)^{1/2}(\rho_{\rm
ej}/\rho_{\rm ext})^{1/4}\right]$. During the relativistic
self-similar stage, the Lorentz factor $\Gamma$ of the shocked
external medium scales as a power law with radius $R$. The scaling is
easy to understand if one keeps in mind that in the (downstream) rest
frame of the shocked fluid the cold upstream fluid comes in with a
bulk Lorentz factor of $\Gamma$, and at the shock front this ordered
bulk motion turns into random motion with the same average Lorentz
factor. In the lab frame the average energy per particle is larger
roughly by a factor of $\Gamma$ compared to the downstream rest frame
of the shocked fluid, and therefore $E \approx \Gamma^2 Mc^2$ where
$M$ is the total amount of swept-up (rest) mass. For a Uniform medium
$M \propto R^3$ and therefore for an adiabatic blast wave (for which
$E \approx {\rm const}$) $\Gamma = \eta(R/R_{\rm dec})^{-3/2}$ since
$\eta = \Gamma(R_{\rm dec})$.  The relativistic self-similar stage
ends at $t_{\rm NR} \approx (3E/4\pi\rho_{\rm ext}c^5)^{1/3} \sim
\eta^{8/3}t_{\rm dec}$ when the flow becomes non-relativistic, after
which it approaches the Newtonian Sedov-Taylor solution.

\subsection{Emission Mechanism}
\label{sec:emission}

\begin{figure}
\rotatebox{270}{
\includegraphics[width=0.74\columnwidth]{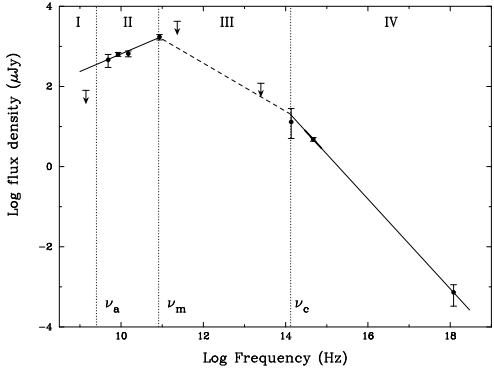}}
\caption{The X-ray to radio afterglow spectrum of GRB~970508 at 12.1
days after the GRB \citep[from][]{Galama98}. A fit to the
low-frequency part ($4.86-86\;$GHz) with $F_\nu \propto \nu^{0.44\pm
0.07}$ is shown as well as the extrapolation from X-ray to optical
({\it solid lines}). The local optical spectral slope ($2.1-5.0\;$days
after the event) is indicated by the thick solid line. Also indicated
is the extrapolation $F_\nu \propto \nu^{-0.6}$ ({\it dashed
line}). Indicated are the rough estimates of the break frequencies
$\nu_a$ (due to self-absorption), $\nu_m$ (typical synchrotron
frequency) and $\nu_c$ (due to electron cooling).}
\label{fig:970508spec}
\end{figure}

The dominant emission mechanism in the afterglow is thought to be
synchrotron radiation, of shock-accelerated relativistic electrons
that gyrate in the post-shock magnetic fields. This is supported by
the broad band afterglow spectrum, which consists of several power law
segments \citep[e.g.,][see Figure~\ref{fig:970508spec}]{Galama98},
and by the detection of linear polarization at the level of a few
percent in the optical or NIR afterglows of several GRBs
\citep{Covino99, Wijers99,Rol00,Covino03}.  Synchrotron self-Compton
(SSC; the inverse-Compton scattering of the synchrotron photons by the
same population of relativistic electrons that emits the synchrotron
photons) can sometimes dominate the afterglow flux in the X-rays
\citep{SE01,Harrison01}. The power-law nature of the broad band
spectrum suggests a power law energy distribution of shock-accelerated
electrons.

\begin{figure}
\includegraphics[width=1.0\columnwidth]{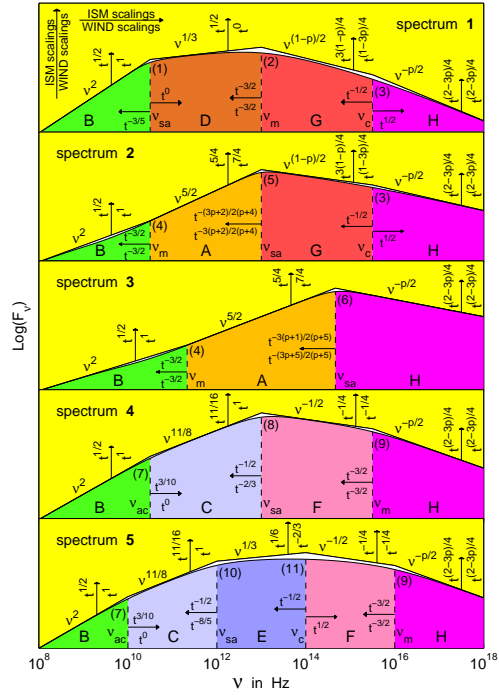}
\caption{The afterglow synchrotron spectrum, calculated for the
\citet{BM76} spherical self-similar solution, under standard
assumptions, using the accurate form of the synchrotron spectral
emissivity and integration over the emission from the whole volume of
shocked material behind the forward (afterglow) shock \citep[for
details see][]{GS02}.  The different panels show the possible broad
band spectra of the afterglow synchrotron emission, each corresponding
to a different ordering of the spectral break frequencies.}
\label{fig:spec}
\end{figure}

The electrons and magnetic fields behind the shock are assumed to hold
some fixed fractions ($\epsilon_e$ and $\epsilon_B$, respectively) of
the internal energy. This reproduces the observed spectrum rather
well, while together with the simple spherical self-similar dynamics
outlined above, it predicts light curves which also consist of several
power law segments. A change in the temporal index occurs either when
a spectral break frequency sweeps past the observed frequency (in
which case it is accompanied by a change in the spectral slope near
the observed frequency), or when a hydrodynamic transition occurs (in
which case no change in the spectral slope is expected across the
temporal break). Figure~\ref{fig:spec} shows the various options for
the broad band spectrum, as well as the temporal scaling of the flux
within each of the different power law segments of the spectrum for a
spherical adiabatic blast wave given by the \citet{BM76} self-similar
solution, for either a uniform external medium, or a wind-like
external density profile (which is discussed in
\S~\ref{sec:variants}).

\subsection{Variants of the Basic Model}
\label{sec:variants}

Several variations on the simplest model have been discussed in the
literature, as well as their observational signatures, many of which
were proposed before their observational manifestations were detected.
Radiative losses have been considered \citep{BM76,CPS98,BD00}, and
result in a faster decrease of the Lorentz factor with radius, since
$\Gamma^2 \approx E(R)/M(R)c^2$, and also with the observed
time. This, in turn, together with the decrease in time of the energy
in the afterglow shock result in a faster decay of the optical and
X-ray fluxes. Conversely, energy injection into the afterglow shock
\citep[e.g.,][]{SM00}, would result in a slower flux decay.

A general power law index for the external density, $\rho_{\rm ext} =
Ar^{-k}$ with $k < 3$ was considered early on \citep{BM76}. It was
only much later, however, when a specific theoretical motivation for
such a non-uniform external density profile in the context of GRBs was
pointed out \citep{CL99}, as evidence accumulated in favor of a
massive star progenitor that is expected to be surrounded by its
pre-supernova stellar wind with $k \approx 2$. A steeper external
density profile (larger $k$) results in a faster flux decay in the
optical (and also in X-rays, if below the cooling frequency $\nu_c$).

It has also been realized that, similar to other astrophysical sources
of relativistic outflow such as active galactic nuclei and
micro-quasars, the GRB outflow is also expected to be collimated into
narrow bipolar jets \citep[e.g.,][]{Rhoads97}. This idea became even
more compelling as the measured redshifts of several GRBs, which
became available thanks to the detection of their afterglows, implied
very large energy output in $\gamma$-rays assuming isotropic emission,
$E_{\rm\gamma,iso}$, which approached and in one case (GRB~991023)
even exceeded a solar rest energy. If most of the $\gamma$-rays are
emitted within a small fraction, $f_b \ll 1$, of the total solid angle
(where $f_b \approx \theta_0^2/2$ for conical uniform narrow bipolar
jets of initial half-opening angle $\theta_0$), then the true energy
output in $\gamma$-rays, $E_\gamma$, is much smaller than its
isotropic equivalent value, $E_\gamma = f_b E_{\rm\gamma,iso}$.

\begin{figure}
\includegraphics[width=1.0\columnwidth]{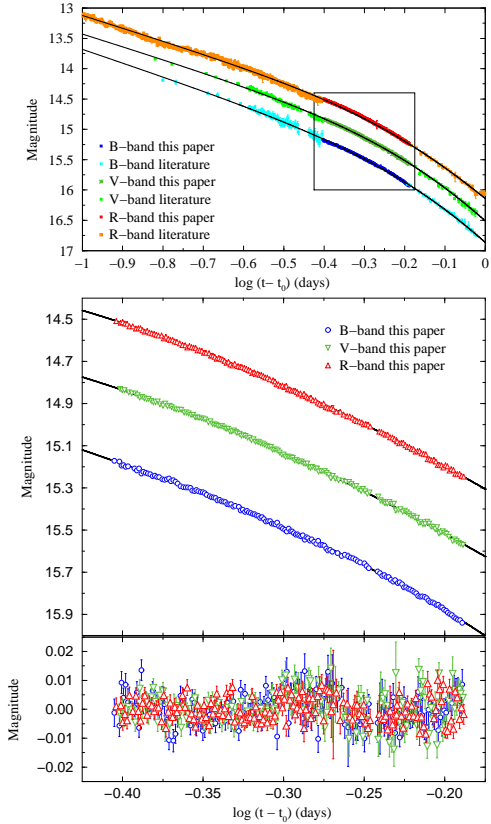}
\caption{Optical light curves at three different frequencies
($BVR$-bands) of the GRB~030329, which show a clear achromatic break,
attributed to a jet, and are fitted to a model featuring two power
laws that join smoothly at the jet break time
\citep[from][]{Gorosabel06}. The bottom panel shows the fit residuals.}
\label{fig:jet-break}
\end{figure}

Simple semi-analytic models for uniform conical jets which expand
sideways rapidly when their Lorentz factor $\Gamma$ drops below
$\theta_0^{-1}$ predicted an achromatic steepening of the afterglow
flux decay, i.e. a ``jet break'' in the afterglow light curve
\citep{Rhoads97,Rhoads99,SPH99}. Such jet breaks in the afterglow
light curves of many GRBs were indeed detected soon thereafter (an
example is shown in Figure~\ref{fig:jet-break}), which provided good
support for GRB outflows being collimated into narrow jets. The
structure and dynamics of GRB jets, however, are still not fully
explored \citep[for a recent review see][]{Granot07a}.

In the pre-{\it Swift} era, while most afterglow light curves showed a
smooth power law decay \citep{Stanek99,LS03,Gorosabel06}, and
sometimes also a ``jet break'', some optical afterglows have shown
significant temporal variability, with strong deviations from the more
typical smooth power law behavior. The best examples are GRBs 021004
\citep{Pandey02,Fox03,Bersier03} and 030329
\citep{Lipkin04}. Possible variants of the
simplest standard afterglow model have been suggested in order to
account for such temporal variability in GRB afterglow light
curves. These include variations in the external density
\citep[e.g.,][see, however,
\citealt{NG07}]{WL00,Lazzati02,NPG03,NP03}, or in the energy of the
afterglow shock. The latter includes energy injection by ``refreshed
shocks'' -- slower shells of ejecta that catch up with the afterglow
shock on long time scales
\citep[e.g.,][]{RM98,KP00a,SM00,R-RMR01,GNP03} or a ``patchy shell''
-- angular inhomogeneities within the outflow
\citep[e.g.,][]{KP00b,NPG03,HP03,NO04}. Another possible cause for
variability in the afterglow light curve, although it is expected to
be quite rare, is microlensing by an intervening star in a galaxy that
happens to be close to our line of sight \citep{GLS00,GGL01}.

\begin{figure}
\includegraphics[width=1.0\columnwidth]{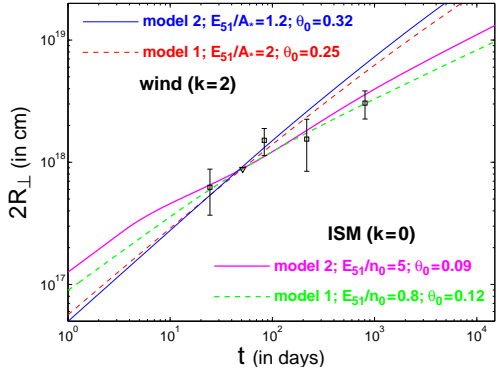}
\caption{ Tentative fits of theoretical models for the evolution of
the source size \citep[from][]{GR-RL05} to the observed image size (of
diameter $2R_\perp$) of the radio afterglow of GRB~030329
\citep[from][]{Pihlstrom07}. In model 1 there is relativistic lateral
spreading of the GRB jet in its local rest frame, while in model 2
there is no significant lateral expansion until the jet becomes
non-relativistic. The external density is taken to be a power law with
the distance $r$ from the source, $\rho_{\rm ext} = Ar^{-k}$, where $k
= 0$ for a uniform external density while $k = 2$ is expected for a
stellar wind environment.}
\label{fig:image}
\end{figure}

Finally, it is worth mentioning another success of the standard
afterglow model: the prediction for the size of the afterglow
image which agrees very well with the available observations. For
GRB~970508 an estimate of the image size of $\sim 10^{17}\;$cm after
$\sim 30\;$days was obtained from the quenching of diffractive
scintillations in the radio \citep{Frail97,WKF98}. For GRB~030329 the
size of the afterglow image was measured directly by radio
interferometry with the VLBA at several different epochs
\citep{Taylor04,Taylor05,Pihlstrom07} and agrees well with the
expectations of the standard afterglow model
\citep{ONP04,GR-RL05}, as illustrated in Figure~\ref{fig:image}.

\section{Some Weaknesses of the Standard Afterglow Model}
\label{sec:weak}

Despite its many successes, the standard afterglow model has several
weaknesses, two of which we briefly mention here.

The study of the physics of relativistic collisionless shocks is still
at its infancy, and the standard afterglow model simply parameterizes
our ignorance. The usual assumptions are that (i) the magnetic field
everywhere within the shocked region holds a constant fraction,
$\epsilon_B$, of the internal energy, and (ii) just behind the shock
front all electrons are accelerated into a power law energy
distribution (of index $p$) with a sharp low energy cutoff and hold a
fixed fraction, $\epsilon_e$, of the internal energy. The validity of
these assumptions is not clear, and they are not fully supported by
first principles calculations. In principle, even if the electron
energy distribution may be reasonably approximated as a power law, the
parameters $\epsilon_e$ and $p$ may vary with the shock Lorentz factor
and the upstream composition. The same holds for $\epsilon_B$, which
may also vary (decrease) with the distance behind the shock. Moreover,
likely only some fraction $\xi_e < 1$ of the electrons are accelerated
into a power-law distribution of energies (the rest forming a
quasi-thermal population). 

This is an active field of research, with a lot of recent progress
involving both analytic work
\citep[e.g.,][]{ML99,KW05,LE06} and particle in cell numerical
simulations \citep[e.g.,][and references
therein]{Spitkovsky07}. Nevertheless, the major questions are still
not fully resolved. These include the decay of the magnetic field
downstream of the shock transition and whether it saturates at some
finite value that can explain afterglow observations within the
standard picture, how particles are acceleration in these shocks and
to what energy distribution.

The dynamics of GRB jets as they sweep-up the external medium,
decelerate, and eventually expand sideways, have still not been
studied in much detail, despite the important implications for GRB
afterglows and GRB physics in general \citep[for a recent review
see][]{Granot07a}. While simple semi-analytic models suggest that
after $\Gamma$ drops below $\theta_0^{-1}$ the jet starts expanding
laterally exponentially with radius \citep{Rhoads99,SPH99}, numerical
simulations show a much more modest degree of lateral expansion where
most of the energy in the flow remains within the original opening
angle as long as the jet is relativistic \citep{Granot01,KG03,CGV04}.
Recently, a self-similar solution has been found \citep{Gruzinov07} in
the lines of the simple semi-analytic models. However, it appears to
be approached very slowly, so that for realistic initial conditions it
may not be fully applicable and the lateral expansion may indeed be
very modest.

\section{The {\it Swift} Era: New Observations and their Implications}
\label{sec:Swift}

\begin{figure}
\rotatebox{270}{
\includegraphics[width=0.77\columnwidth]{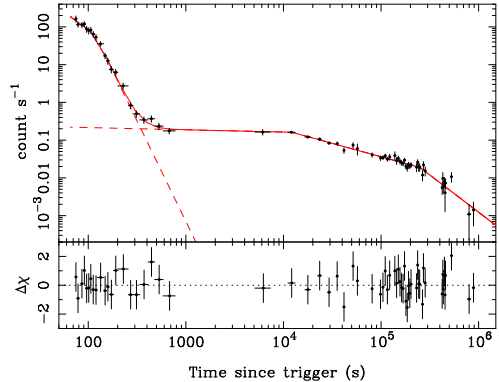}
}
\caption{Light curve of GRB~050315 in the $0.2-5\;$keV band, taken by
the {\it Swift} X-ray telescope (XRT), also showing the best-fitting
model comprising a singly broken power law and a double broken power
law, which dominate at early and late times, respectively
\citep[from][]{Vaughan06}. The lower panel shows the residuals of the
fit.}
\label{fig:050315}
\end{figure}

{\it Swift} has discovered surprising new features in the early X-ray
afterglow \citep{Nousek06}. These included mainly (i) an initial rapid
decay phase where $F_\nu \propto t^{-\alpha}$ with $3 \lesssim
\alpha_1 \lesssim 5$ lasting from the end of the prompt emission up to
$\sim 10^{2.5}\;$s, (ii) a subsequent shallow decay phase where $0.2
\lesssim \alpha_2 \lesssim 0.8$, lasting up to $\sim 10^4\;$s
(followed by the familiar pre-{\it swift} power law decay with $1
\lesssim \alpha_3 \lesssim 1.5$), and (iii) X-ray flares, which appear
to be overlaid on top of the underlying power law decay in stages (i)
and (ii). A good example of such an X-ray light curve that shows these
different stages (but no X-ray flares) as well as a possible jet break,
is GRB~050315, which is shown in Figure~\ref{fig:050315}. The initial
rapid decay stage seems to be a smooth extension of the prompt
emission \citep{Obrien06}, and is therefore most likely the tail of
the prompt GRB, probably due to emission from large angles relative to
our line of sight \citep{KP00}.

The X-ray flares appear to be a distinct emission component, as
suggested by their generally different spectrum compared to the
underlying power law component, and by the fact that the flux after a
flare is usually the continuation of the same underlying power law
component from before the flare \citep[e.g.,][]{Burrows05}. In many
cases these flares show sharp large amplitude flux variation on time
scales $\Delta t \ll t$ \citep[see, e.g.][]{Krimm07}, which are very
hard to produce by the external shock, and suggest a sporadic late
time activity of the central source. An alternative explanation, which
does not require a prolonged central source activity, is delayed
magnetic reconnection events in the outflow (Giannios 2006).

\begin{figure}
\includegraphics[width=1.0\columnwidth]{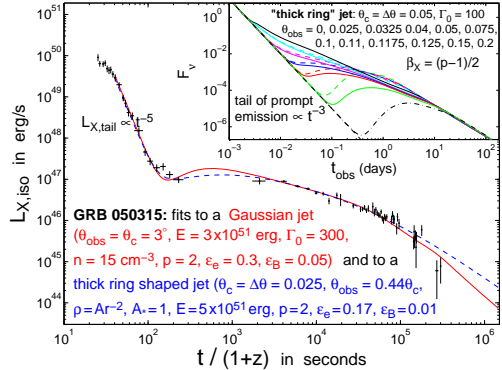}
\caption{A tentative fit to the X-ray light curve of GRB~050315, with
  (i) a Gaussian jet ({\it solid red line}), and (ii) a ring shaped
  (or hollow cone) jet, uniform within
  $\theta_c<\theta<\theta_c+\Delta\theta$ ({\it blue dashed line})
  where in both cases the viewing angle plays a major role in
  producing the shallow decay phase \citep[from][]{EG06}. The initial
  fast decay is attributed to the tail of the prompt emission and
  modeled as a power law $\propto t^{-5}$. The inset shows afterglow
  light curves for a ring shaped jet \citep{Granot05}, for different
  viewing angles $\theta_{\rm obs}$ from the jet symmetry axis.}
\label{fig:viewing-angle}
\end{figure}

\begin{figure}
\includegraphics[width=1.0\columnwidth]{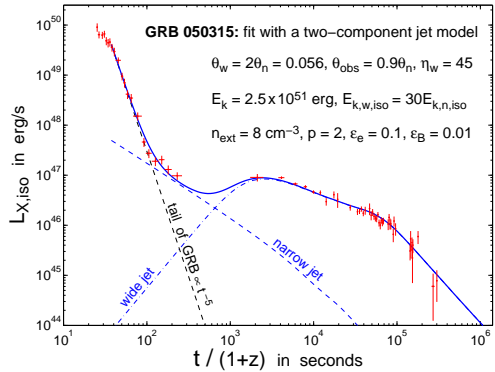}
\caption{Tentative fit to the X-ray light curve of GRB~050315 with the
two-component jet model \citep[from][]{GKP06}. In addition to the
total light curve ({\it thick solid line}) also shown are the separate
contributions of the different components: the tail of the prompt
emission ($\propto t^{-5}$), the narrow outflow, and the wide
outflow. Here $E_k = E_{k,w} + E_{k,n}$ is the total kinetic energy of
the two components. The narrow and wide components occupy the
non-overlapping ranges in the polar angle $\theta$ (measured from the
symmetry axis): $\theta < \theta_n$ and $\theta_n < \theta <
\theta_w$, respectively; $\theta_{\rm obs}$ is the viewing angle with
respect to the jet symmetry axis.}
\label{fig:2comp-jet}
\end{figure}

The shallow decay phase, stage (ii), and the initial rapid decay
phase, stage (i), appear to arise from two physically distinct
emission regions. This is supported by a change in the spectral index
that is observed in some of the transitions between these two stages
\citep{Nousek06}. Furthermore, the shallow decay phase eventually
smoothly steepens into the familiar pre-{\it Swift} flux decay, which
is well established to be afterglow emission (that is attributed to
the forward shock in the standard scenario), strongly suggesting that
the shallow decay phase is similarly afterglow emission. This is also
supported by the fact that there is no evidence for a change in the
spectral index across this break \citep{Nousek06}. There are several
suggestions for the cause of the shallow decay phase, including energy
injection into the afterglow shock (of two main types), viewing angles
slightly outside the region of bright afterglow emission (see
Figure~\ref{fig:viewing-angle}), a complex angular structure of the
jet consisting of two or more components (see
Figure~\ref{fig:2comp-jet}), and time varying shock microphysics
parameters \citep[for a review see][]{Granot07b}. Nevertheless, it is
not clear which of these, if any, is indeed the dominant cause.

The shallow decay phase has interesting implications for GRB and
afterglow theory. It implies a very high efficiency ($\gtrsim 90\%$)
of the prompt $\gamma$-ray emission, unless either the energy in the
afterglow shock at late times ($\gtrsim 10\;$hr) has been
underestimated, or the shock microphysics parameters significantly
vary with time during the early afterglow \citep{GKP06}. Both are
required in order for the $\gamma$-ray efficiency to be $\lesssim
10\%$, which the internal shocks model can reasonably produce. This in
turn requires a large energy in the afterglow shock (typically $\gtrsim
10^{52}\;$erg). Nevertheless, it can still be accommodated within the
framework of the standard afterglow model, albeit, with some
modifications.

Another interesting finding by {\it Swift} is that the early optical
emission, which has been attributed in some cases before {\it Swift} to
the reverse shock, is typically much dimmer than expected
\citep{Roming06}.  This may be since the expectations were too high
\citep{NP04} and motivated by early pre-{\it Swift} detections that
were exceptionally bright. Alternatively, it could be (at least
partly) due to a suppression of the reverse shock in strongly
magnetized GRB outflows \citep{ZK05}, or since {\it Swift} detects
dimmer events (on average) compared to previous instruments, due to
its higher sensitivity. The latter may also partly explain the
relative paucity of detected jet breaks in the {\it Swift} era
\citep{BR07,Sato07,KB08}, although it is not clear yet if this can 
account for all of the effect.

\section{Crisis, Ideas for Solving it, and Future Prospects}
\label{sec:future}

\begin{figure}
\includegraphics[width=1.0\columnwidth]{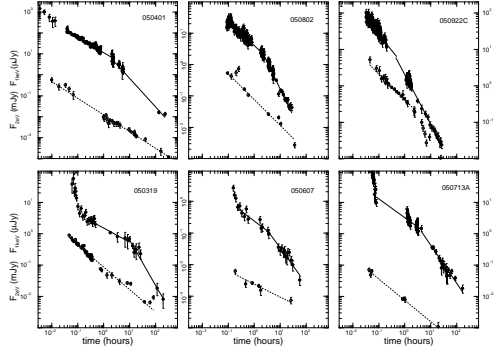}
\caption{Light-curves of six Swift GRB afterglows showing a chromatic
X-ray break that is not seen in the optical at the same time
\citep[from][]{Panaitescu06}. Optical data are shown with open symbols
and are fit with a power-law decay ({\it dotted lines}); X-ray data are
shown with filled symbols and fit with a broken power-law ({\it solid
lines}).}
\label{fig:chromatic}
\end{figure}

Among the new observational features in the {\it Swift} era, the most
difficult one to naturally explain within the framework of the
standard afterglow model is chromatic breaks in the afterglow light
curves. Several GRBs show this feature, where the X-rays show a clear
break (steepening of the flux decay rate) while the optical does not
\citep[e.g.,][see Figure~\ref{fig:chromatic}]{Panaitescu06}. The break
in the X-ray light curve is usually identified with the end of the
shallow decay stage. The optical light curve follows a single power
law decay, usually with a temporal decay index intermediate between
those in the X-rays before and after the break. Such chromatic breaks
appear to be common among the best monitored {\it Swift}
afterglows. Explaining these chromatic breaks requires significant
changes to the existing afterglow model. This marks a crisis for
standard afterglow theory.

Accommodating such chromatic breaks in a model where the X-ray and
optical emissions arise from the same physical region
\citep[e.g.,][]{Panaitescu06} requires not only time dependent
microphysical parameters, but also fine tuning their time (or shock
Lorentz factor) dependence in such a way that will produce a break in
the X-ray but not in the optical, which appears somewhat
contrived. The X-ray and optical may arise in separate physical
components, which would naturally account for their seemingly
decoupled light curves. However, this type of explanation also
introduces a new ingredient (with its associated degrees of freedom)
into the model, and unless it can be tested against other independent
data it is hard to constrain such an option or reach any definitive
conclusions.  One might ask: are we just adding epicycles to a
fundamentally flawed model?

This crisis has lead several people to consider more radical
solutions.  A good example, is the possibility that the afterglow is
dominated by emission from a long-lived reverse shock, as opposed to
emission from the forward shock as in the standard afterglow model
\citep{GDM07,UB07}. This has the advantages of avoiding the need to
produce near equipartition values of the shock microphysics parameters
($\epsilon_B \gtrsim 0.01$ and $\epsilon_e \gtrsim 0.1$) in the
forward shock (which appears to be difficult, and relies on the poorly
understood physics of relativistic collisionless shocks) and reducing
the required efficiency of the prompt $\gamma$-ray emission. On the
other hand, it has its own difficulties, such as tending to
over-produce the observed optical emission (when normalized to the
observed X-ray flux), and a self-absorption frequency that is
typically above the radio band (which is hard to reconcile with radio
afterglow detections, although these are admittedly rare in the {\it
Swift} era).

Moreover, if taken to the extreme, where the emission is dominated by
the reverse shock up to very late times \citep{UB07}, then the
following problem arises. In a few cases the radio afterglows are
observed out to very late times when the flow becomes sub-relativistic
\citep[e.g.,][]{Pihlstrom07} with no apparent transition from being
dominated by one component (the reverse shock) to another (the forward
shock). In the Newtonian regime, however, the properties of GRB remnants
should approach those of supernova remnants, which are often directly
resolved with the radio emission being dominated by the forward shock.

Another rather radical suggestion is that the X-ray emission from the
onset of the shallow decay phase and onwards is dominated by ``late
prompt'' emission \citep{Ghisellini07}, due to prolonged activity of
the central source, while the optical emission during the same time is
dominated by the forward shock emission (i.e. the traditional
afterglow emission).  Since the X-ray and optical emissions in this
model come from distinct physical regions, they are naturally
decoupled. What appears to be less natural, however, is that both the
prompt emission and the ``late prompt'' emission arise from similar
activity of the central source but have very different temporal (very
variable versus smooth) and spectral properties. Furthermore, this
scenario requires an even higher efficiency of the prompt $\gamma$-ray
emission, since the X-ray afterglow flux in this model is below the
observed X-ray flux, implying a smaller energy in the afterglow
shock. It also requires cold acceleration (not driven by thermal
pressure) since after the break in the X-rays the flow Lorentz factor
exceeds the inverse of its half-opening angle.

In summary, as the afterglow observations improve, a more complex
behavior is revealed, which requires the introduction of new
ingredients into the model. Furthermore, some of the basic underlying
physics, such as a detailed study of relativistic collisionless shocks
and the dynamics of GRB jets, are still not fully understood and
require further work. Out of the new observations in the {\it Swift}
era, the most puzzling for afterglow theory are the shallow decay phase
and the chromatic breaks. The former has too many possible
explanations, and it is hard to tell which of them if any is indeed
the dominant cause, while the latter is still awaiting a natural and
convincing explanation without any major problems.
These new observations challenge traditional afterglow 
theory and call for new ideas.

Broad band multi-frequency observations can help constrain the cause
of the shallow decay phase (e.g. distinguish between a highly
relativistic and mildly relativistic long lived reverse shock in the
energy injection scenario) and chromatic breaks in GRB afterglows.
Such observations may also help study the cause of the chromatic
breaks. In particular, observations with the Large Area Telescope
(LAT) on board the Gamma-ray Large Area Space Telescope (GLAST, to be
launched in early 2008) may give us a better handle on the SSC
component which would provide more constraints on the model, and help
distinguish between different scenarios.

\vspace{0.55cm}

The author gratefully acknowledges a Royal Society Wolfson Research
Merit Award.

\end{document}